\documentclass[atoms,conferenceproceedings,submit,moreauthors,pdftex,10pt,a4paper]{mdpi} 

\firstpage{1} 
\articlenumber{x}
\doinum{10.3390/------}
\pubvolume{xx}
\pubyear{2017}
\copyrightyear{2017}
\externaleditor{Academic Editor: name}
\history{Received: September 10, 2017; Accepted: October 23, 2017; Published: date}

\Title{MODELS OF EMISSION LINE PROFILES AND SPECTRAL ENERGY DISTRIBUTIONS TO CHARACTERIZE THE MULTI-FREQUENCY PROPERTIES OF ACTIVE GALACTIC NUCLEI}

\Author{Giovanni La Mura$^{1,\dagger,}$*\orcidA{}, Marco Berton$^{1,2,\ddagger}$, Sina Chen$^{1,\ddagger}$, Abhishek Chougule$^{1,\ddagger}$, Stefano Ciroi$^{1,\ddagger}$, Enrico Congiu$^{1,2,\ddagger}$, Valentina Cracco$^{1,\ddagger}$, Michele Frezzato$^{1,\ddagger}$, Sabrina Mordini$^{1,\ddagger}$ and Piero Rafanelli$^{1,\ddagger}$}

\AuthorNames{Giovanni La Mura, Marco Berton, Sina Chen, Abhishek Chougule, Stefano Ciroi, Enrico Congiu, Valentina Cracco, Michele Frezzato, Sabrina Mordini and Piero Rafanelli}

\address{%
$^{1}$ \quad Department of Physics and Astronomy - University of Padua, Vicolo dell'Osservatorio 3, 35122 - Padova (ITALY) \\
$^{2}$ \quad Astronomical Observatory of Brera - National Institute of Astrophysics (INAF), Via Bianchi 46, 23807, Merate (ITALY)}

\corres{Correspondence: giovanni.lamura@unipd.it; Tel.: +39 (0)49-827-8235}

\firstnote{Current address: Vicolo dell'Osservatorio 3, 35122 - Padova (ITALY)} 
\secondnote{These authors contributed equally to this work.}

\abstract{The spectra of Active Galactic Nuclei (AGNs) are often characterized by a wealth of emission lines with different profiles and intensity ratios that led to a complicated classification. Their electro-magnetic radiation spans more than 10 orders of magnitude in frequency. In spite of the differences between various classes, the origin of their activity is attributed to a combination of emitting components, surrounding an accreting Super Massive Black Hole, in the so called Unified Model. Currently, the execution of sky surveys, with instruments operating at various frequencies, provides the possibility to detect and to investigate the properties of AGNs on very large statistical samples. Thanks to the spectroscopic surveys that allow investigation of many objects, we have the opportunity to place new constraints on the nature and evolution of AGNs. In this contribution we present the results obtained by working on multi-frequency data and we discuss their relations with the available optical spectra. We compare our findings with the AGN Unified Model predictions, and we present a revised technique to select AGNs of different types from other line emitting objects. We discuss the multi-frequency properties in terms of the innermost structures of the sources.}


\conferencetitle{11$^{\rm th}$ Serbian Conference on Spectral Line Shapes in Astrophysics}

\begin{document}
\newcommand{\ion}[2]{#1~{\small #2}}


\section{Introduction}

When talking about Active Galactic Nuclei (AGNs), we generally refer to the nuclei of galaxies where a Super Massive Black Hole (SMBH), with a mass ranging between $10^6\,{\rm M}_\odot$ and $10^{10}\,{\rm M}_\odot$, is fed by a continuous flow of matter from the surrounding environment. This process, denoted as \emph{accretion}, leads to the conversion of the gravitational binding energy of the accreted material into heat and radiative energy, through the effects of the viscous interactions that arise in the accreted matter, as  it is accelerated up to several thousands ${\rm km\, s}^{-1}$ by the strong gravitational pull of the black hole \cite{Blandford86}. In spite of this common interpretation, AGNs present a wide range of striking observational differences in their spectra, in their total power and in the frequency range where most of their energy is radiated away. While the first ones to be clearly identified were located in galaxies with exceptionally bright optical nuclei \cite{Seyfert43}, nearly 10\% of the total population were subsequently found to radiate large fractions of their power in the radio and the high-energy domain \cite{Elvis94}. Even their total luminosities can change over a wide range, which is typically considered to lie between $10^{40}\,{\rm erg\, s}^{-1}$, in the case of low-luminosity sources, up to some $10^{46}\,{\rm erg\, s}^{-1}$ for the most powerful ones. They are distributed from the local Universe, where the low-luminosity objects are more common, all the way up to very high redshifts ($z \geq 7$), where powerful activity becomes more frequent.

Looking at the characteristics of the optical-UV spectra, AGNs are generally characterized by the presence of a non-thermal continuum, often well represented by a power-law shaped spectrum of the form $L_\nu \propto \nu^{-\alpha}$, sometimes accompanied by prominent emission lines with different profiles. In some cases, we additionally observe a hot thermal excess, with a peak that likely falls in the far UV, or different contributions from the host galaxy stellar populations. Following a scheme that was first outlined by \cite{Khachikian74}, we generally call Type 1 AGNs those objects whose spectra show broad recombination lines, with profiles corresponding to velocity fields exceding $1000\, {\rm km\, s}^{-1}$ in Full Width at Half Maximum (FWHM), from H, He, or from other permitted lines of heavy ions like \ion{Fe}{II}, \ion{C}{IV} and \ion{Si}{IV}, together with narrow forbidden lines (FWHM$\approx 300 - 500\, {\rm km\, s}^{-1}$) from e.g. [\ion{Ne}{V}], [\ion{Ne}{III}], [\ion{O}{III}], [\ion{O}{II}], [\ion{O}{I}], [\ion{N}{II}] and [\ion{S}{II}]. We call instead Type 2 AGNs those where both permitted and forbidden lines only have narrow profiles. In general, it is observed that Type 1 sources are brighter and they commonly show a thermal UV excess, while Type 2 objects are dimmer and more severly contaminated by the host galaxy spectral contributions.

The most widely accepted way to explain the observations is to assume that the central accreting SMBH is surrounded by a hot accretion disk, radiating in the optical, UV and X-Ray frequencies, and a compact region (less than $0.1\,$pc in size) of dense ionized gas ($N_e \geq 10^9\, {\rm cm}^{-3}$), which produces Doppler broadened recombination lines, due to the large gravitational acceleration, and is therefore called the {\em Broad Line Region} (BLR). The gas that is located at larger distances ($1\,{\rm pc} \leq r \leq 1\,{\rm kpc}$), though being still ionized and producing emission lines, has a considerably smaller velocity field and electron densities closer to typical nebular environments ($10^3\, {\rm cm}^{-3} \leq N_e \leq 10^6\, {\rm cm}^{-3}$). It, therefore, radiates both permitted and forbidden lines with narrow Doppler profiles, giving raise to what we call the {\em Narrow Line Region} (NLR). If the central structure is partially obscured by an optically thick distribution of matter, as it is supported by some observational evidence \cite{Antonucci85}, the difference between Type 1 and Type 2 objects is consistently explained by the fact that our line of sight, respectively, may or may not reach the central regions, without being intercepted by the surrounding material, in what is called the AGN \emph{Unified Model} \cite{Antonucci93}. When the accretion flow becomes coupled with the magnetic fields that develop close to the SMBH in such a way that a relativistically beamed jet of plasma is accelerated away from the nucleus, AGNs become powerful sources of radio and high-energy emission, eventually developing extended radio morphologies \cite{Urry95}.

In spite of the fairly comprehensive interpretation, AGNs still pose many fundamental questions, because most of the relevant physical processes involved in the accretion and in the acceleration of jets are confined close to the SMBH, in a region that lies still beyond the resolution capabilities of present-day instruments. For this reason, a large part of our current knolwedge concerning AGNs, is based on the analysis of their spectra and on monitoring of the correlations existing among their numerous spectral components. However, not all sources are equally good for such investigations, since Type 1 objects tend to be dominated by the emission of the AGN, while Type 2 sources are strongly affected by obscuration and stellar light contributions from the host galaxy. The main details to understand the Physics of AGN, therefore, can only be constrained by careful inspection of spectral properties that should be possibly extended over large statistical samples, in order to compare the predictions of different models with the available observations. In recent times, several campaigns have been carried out to monitor the sky at different frequencies and to obtain spectroscopic observations. In this contribution, we describe a revised scheme to select AGNs based on the properties of their emission lines and colors, we illustrate the potential of multi-frequency models to relate the observed Spectral Energy Distributions (SEDs) with optical spectra, and we discuss the information that we are able to extract on their central structures from the combined analysis of line profiles and multi-frequency data.

\section{Results}

The most common property shared by different types of AGNs is the emission of an intense continuum of non-thermal radiation that can possibly ionize diffuse gas and, thus, give raise to emission lines in the spectra. Since lines excited by non-thermal ionizing continua differ in intensity and distribution from lines excited by the continuum of hot thermal sources \cite{Baldwin81, Veilleux87}, we can apply a set of diagnostic diagrams, based on the intensity ratios of specific lines, in order to recognize the footprint of ionization from thermal and non-thermal sources in external galaxies. When the amount of spectroscopic data was dramtically increased, thanks to the execution of large spectroscopic surveys, like the \emph{Sloan Digital Sky Survey} \cite{York00, Blanton17}, this method was further refined, demonstrating its ability to recognize the presence of obscured AGN activity \cite{Kewley06}. This kind of information is fundamental to assess the statistical relations existing between obscured and unobscured sources, which constrain the structure of the central source and its possible dependence on luminosity or age. In order to perform such a study, however, we need an instrument that is in principle able to detect different types of AGN activity, with the smallest possible influence of selection effects. We obtained such a tool by combining spectroscopic and photometric parameters, based on an investigation of how different spectral classes are related with multi-frequency emission.

\begin{figure}[t]
  \begin{center}
    \includegraphics[width=0.32\textwidth]{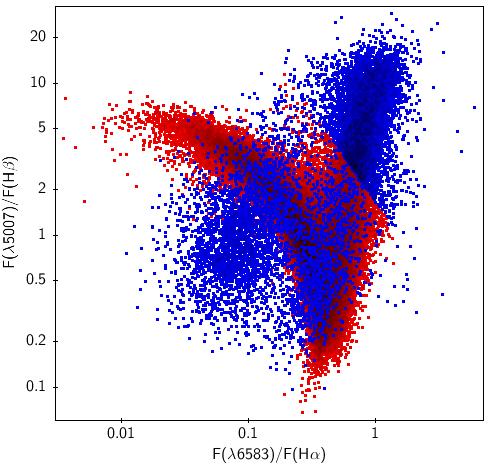}
    \includegraphics[width=0.32\textwidth]{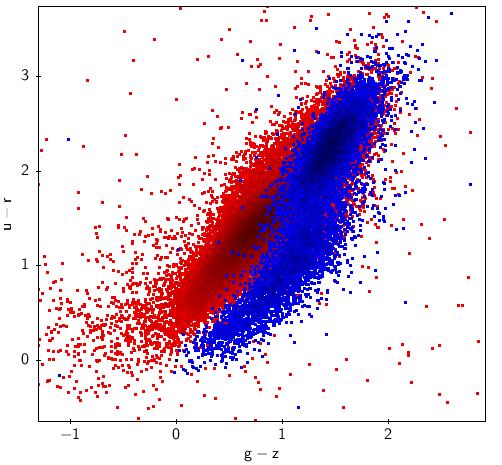}
    \includegraphics[width=0.32\textwidth]{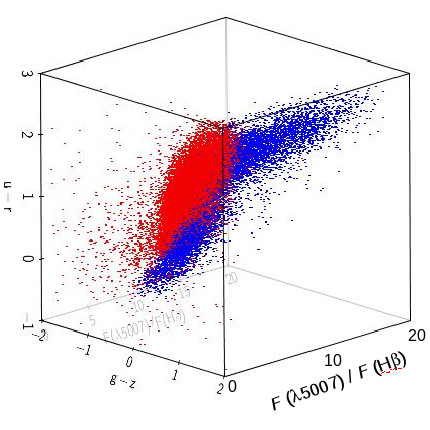}
  \end{center}
  \caption{Distribution of AGN (blue points) and star formation (red points) powered objects with emission line spectra on the [\ion{O}{III}]$\lambda 5007$/H$\beta$\ vs. [\ion{N}{II}]$\lambda 6583$/H$\alpha$\ diagnostic diagram (left panel), on the $u - r$ vs. $g - z$ color-color diagram (middle panel) and on the three dimensional parameter space combining the two-color photometry with the [\ion{O}{III}]$\lambda 5007$/H$\beta$\ diagnostic ratio. The distribution in the three dimensional parameter space illustrates how Type 1 AGNs are best identified among line emitting sources by means of photometric criteria, while Type 2 sources can be distinguished on the basis of their spectral properties.}
\end{figure}
\subsection{AGN selection from spectroscopic surveys}

The investigation of AGN statistical properties requires to revise the selection techniques, which have been classically designed to detect specific types of sources, based on their characteristic properties. While Type 1 AGNs are generally well identified by the presence of prominent broad emission lines in the spectra, which are commonly accompanied by a hot thermal continuum excess, some times referred to as the \emph{Big Blue Bump}, Type 2 AGNs are only characterized by narrow emission lines, which can also be present in the spectra of galaxies with strong star formation activity. In the case of narrow line emitting sources, the methods based on diagnostic diagrams are fairly well suited to distinguish between AGN and stellar activity, but if we perform a selection of spectra simply based on the presence of emission lines, to collect different types of sources, the emission line diagnostics alone are not straightforwardly applicable. An example of this effect is shown in Fig. 1, where we compare different methods to distinguish AGNs from thermally excited spectral line emitters. The reason for which the classic diagnostic ratios cannot be used on a sample of objects including Type 1 sources resides in the use of the recombination lines, which are needed to normalize the strength of the forbidden lines that probe the temperature and the ionization structure of the gas. The presence of a strong contribution in the broad component of the recombination lines in Type 1 objects, which is not balanced in the forbidden lines, forces Type 1 sources to populate the non-AGN region of the plots. This is obvious, because this type of diagnostic diagram is designed to work on the emission of the NLR alone and cannot account for the BLR component. It has been proposed that a different choice of the emission line diagnostic ratios, involving only forbidden lines, might in principle solve the problem \cite{Vaona12}, but the available choices either involve the use of weak lines, or they are strongly subject to the effects of interstellar extinction.

If the direct sight of the central regions in Type 1 AGNs can give raise to problems in recognizing their spectral signature, on the other hand we have the opportunity to take advantage from the strong blue and UV continuum, which is produced by the central source and is not obscured along the line of sight. It has been shown that Type 1 AGNs can effectively be selected by means of photometric criteria that compare their colors with those of non-active objects, to the extent that the SDSS uses a photometric pipeline to select QUASAR candidates for follow-up spectroscopy \cite{Richards02, Schneider10}. This method has some limits when comparing objects at very different redshifts, but, in the low redshift domain where the diagnostic lines are still available in the optical frequency window ($z \leq 0.5$), it defines a well-established parameter space where Type 1 AGNs can be effectively distinguished from other line emitting sources. A projection of this parameter space on the $u - r$ vs. $g - z$ color-color diagram is also illustrated in Fig.~1. The choice of this extended color bands, that was based on the extinction corrected magnitude measurments of all SDSS objects with emission lines of H$\alpha$, H$\beta$, [\ion{O}{III}]$\lambda 5007$ and [\ion{N}{II}]$\lambda 6583$ datected at more than $5 \sigma$\ level, maximizes the effect of the blue continuum of Type 1 sources, over the stellar continuum of other sources. As a consequence, we recover the possibility of detect different types of nuclear activity, by combining classic spectroscopic diagnostics and photometric colors into a multi-dimensional parameter space, where AGNs populate a separate sequence with respect to other non-AGN powered sources.

\subsection{Emission lines and models of AGN Spectral Energy Distributions}

The selection of general samples of AGNs belonging to different spectral classes allow to search for observations of the corresponding sources in multiple frequencies. By combining the available data, it is possible to reconstruct the AGN SEDs and to compare them with the associated optical spectra, as it is illustrated for example in Fig.~2. The plots show the different SEDs of two prototypical AGNs (3C~273 for Type 1 and NGC~1068 for Type 2) modelled through a combination of thermal and non-thermal radiation components, together with their characteristic spectra. We can immediately appreciate how the occurrence of a Type 1 spectrum, with broad lines and blue continuum, is well associated with a strong dominance of the non-thermal contribution and a direct sight towards the hottest central regions, resulting in an excess of ionizing radiation, in agreement with the Unified Model predictions. Conversely, the Type 2 SED is totally consistent with an obscured central source, whose low energy ionizing radiation is severely absorbed and reprocessed by a distribution of material that subsequently re-emits photons in the IR domain. Only the more penetrating high-energy photons and the long wavelength radio emission, which is practically unaffected by obscuration, can propagate directly from the central source, therefore resulting in a optical spectrum that is dominated by the host galaxy and the emission lines coming from the unobscured NLR.

\begin{figure}[t]
  \begin{center}
    \includegraphics[width=0.46\textwidth]{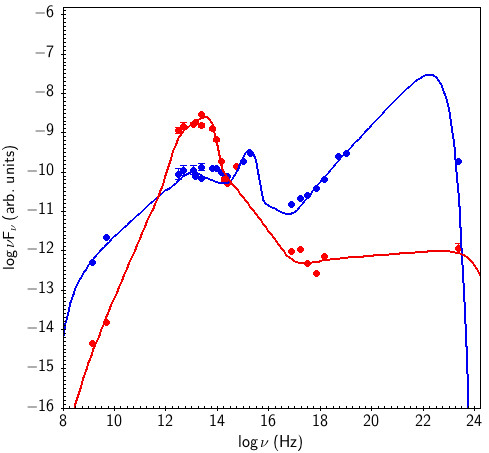}
    \includegraphics[width=0.49\textwidth]{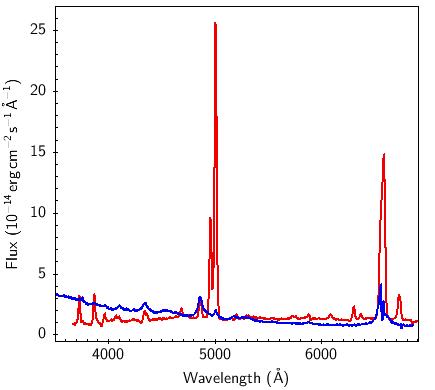}
  \end{center}
  \caption{Examples of SED models (left panel) and optical spectra (right panel) for the prototypical Type 1 AGN 3C 273 (blue line and points) and the prototypical Type 2 AGN NGC 1068 (red line and points). The SEDs have been normalized to the same optical V band magnitude and the models are based on the combination of two cut-off power laws and one black body contribution. The spectrum of 3C 273 has been observed at the Asiago Astrophysical Observatory, while the spectrum of NGC 1068 comes from literature \cite{Moustakas06}. Both spectra were taken to rest frame wavelength scale.}
\end{figure}
In addition to provide observational evidence in support to the Unified Model, and possibly to associate different degrees of absorption and light reprocessing to various AGN classes, the combination of SED models and spectroscopic observations is a promising instrument to improve our knowledge of AGN distribution. Thanks to the numerous efforts that have been devoted to surveying the whole sky at various frequencies, it is now possible to recognize AGN candidates from their SED, even in sky areas that have not yet been covered by detailed and publicly available spectroscopic programs. The further possibility that models of the observed SED may lead to a prediction of the expected AGN class, in addition to optimizing the execution of follow-up campaigns, provides statistical constraints to infer structural and evolutionary details in large samples of AGNs.

\subsection{Multi-frequency analysis of the central engine}

Due to the extremely compact size of the regions where the continuum and the bulk of emission lines are produced in AGNs, we do not yet have a fully developed interpretation of their innermost structures. Most of our current understanding derives from the analysis of spectra and from models that carry out the inferred physical conditions. Therefore, the extension of spectroscopic analysis to different frequencies improves our ability to explore unresolved structures. Such an example is illustrated in Fig.~3 for the case of the QSO PG~1114+445, where we compare the \emph{XMM Newton} X-ray spectrum with measurements of the normalized intensities of the broad components of the Balmer lines of Hydrogen. This quantity is defined as:
\begin{equation}
  I_n = \dfrac{\lambda_{ul} I_{ul}}{g_u A_{ul}},
\end{equation}
where $\lambda_{ul}$ is the wavelength corresponding to the transition from an upper level $u$ to a lower level $l$ (with $l = 2$ for H Balmer lines), $I_{ul}$ the measured intensity, $g_u$ the statistical weight of the upper level and $A_{ul}$ the spontaneous transition probability. In the case of an optically thin line, we can use the general expression:
\begin{equation}
  I_{ul} = \dfrac{1}{4 \pi} \dfrac{h c}{\lambda_{ul}} A_{ul} \int_0^{s^*} N_u {\rm d}s,
\end{equation}
where $h$ and $c$ are the Planck's constant and the vacuum speed of light, while $N_u$ is the concentration of atoms in the upper level and the integration is executed throughout the extension of the source, to obtain:
\begin{equation}
  I_n = \dfrac{1}{4 \pi} \dfrac{h c}{g_u} \int_0^{s^*} N_u {\rm d}s.
\end{equation}
If we now assume that the distribution of emitting atoms is spatially constant within the source and that it can be represented by a Boltzmann formula at least in the high excitation stages (a condition known as \emph{Partial Local Thermodynamic Equilibrium}, PLTE), we obtain:
\begin{equation}
  I_n = \dfrac{1}{4 \pi} \dfrac{h c}{g_1} N_1 s^* \exp \left( - \dfrac{\Delta E_u}{k_B T_e} \right),
\end{equation}
where $\Delta E_u$ is the upper level excitation energy. It is therefore clear that, in the assumed conditions, we expect that:
\begin{equation}
  \log I_n = \log \left( \dfrac{1}{4 \pi} \dfrac{h c}{g_1} N_1 s^* \right) - \dfrac{\log e}{k_B T_e} \Delta E_u,
\end{equation}
which is a linear function of the upper level's excitation energy \cite{Popovic03}.

\begin{figure}[t]
  \begin{center}
    \includegraphics[width=0.48\textwidth]{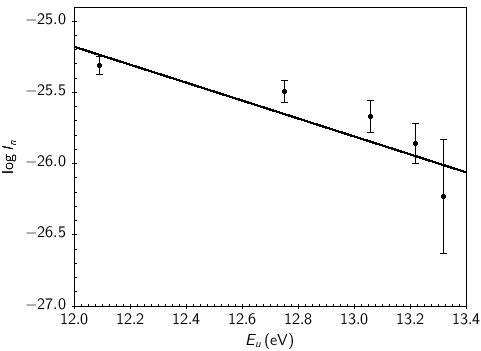}
    \includegraphics[width=0.48\textwidth]{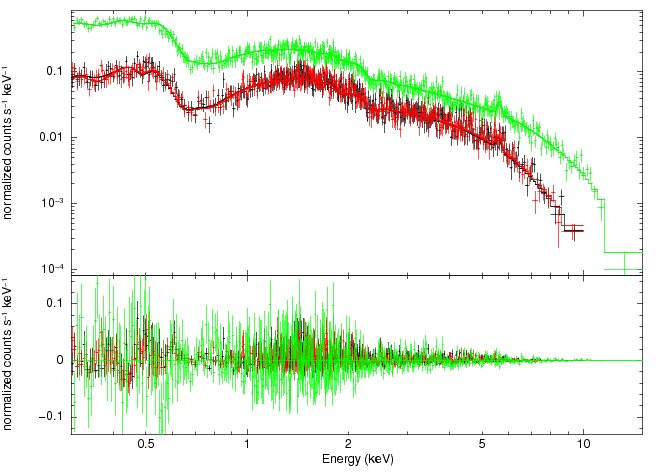}
  \end{center}
  \caption{Suppression of high order Balmer lines, observed through the normalized intensities of the broad components in the optical spectrum of PG~1114+445 (left panel), compared with the soft X-ray spectrum obtained from the \emph{XMM Newton} EPIC instrument (green points) and from the twin MOS cameras (red and black points). The X-ray spectrum is modelled with a power-law spectrum, a soxt X-ray thermal excess and an ionized absorber. Absorption from the neutral medium within the Milky Way is also taken into account.}
\end{figure}
The deviation from the expected linear behavior observed in Fig.~3 is an indication that the flux of high order Balmer line photons is lower than the prediction, which can happen in the presence of a dense layer of recombining plasma. In this case, indeed, Eq.~(2) would be modified in:
\begin{equation}
  I_{ul} = \dfrac{1}{4 \pi} \dfrac{h c}{\lambda_{ul}} A_{ul} \int_0^{s^*} N_u e^{-k_{lu} s} {\rm d}s,
\end{equation}
where
\begin{equation}
  k_{lu} = N_l \sigma_{lu}
\end{equation}
is the line absorption coefficient, controlled by the density of ions in the lower level $N_l$ and the line photon absorption cross-section $\sigma_{lu}$. In typical nebular conditions, it would be $k_{lu} << 1$ for any $l > 1$, but, in presence of a thick layer of ionized plasma, like the one responsible for the observed absorption in the soft X-ray spectrum, which has an estimated column density $N_C = 4.896 \cdot 10^{21}\, {\rm cm}^{-2}$, it could be no longer negligible. In such circumstances, indeed, we expect the lower level of the Balmer series to be overpopulated by recombination processes and by Ly$\alpha$\ photon trapping. If, therefore, this X-ray absorber is located outside the BLR, it could very likely be responsible for the absorption of Balmer photons, as well.

\begin{figure}[t]
  \begin{center}
    \includegraphics[width=0.45\textwidth]{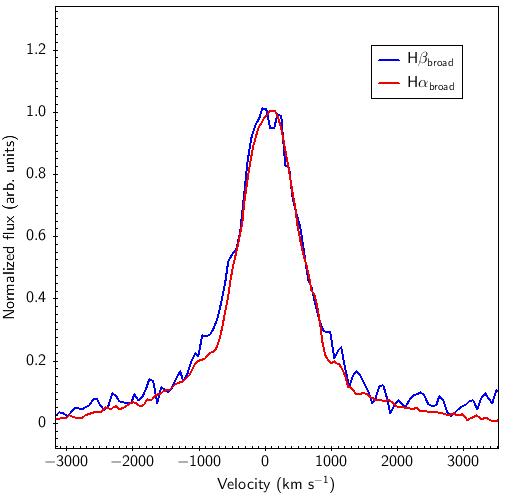}
    \includegraphics[width=0.48\textwidth]{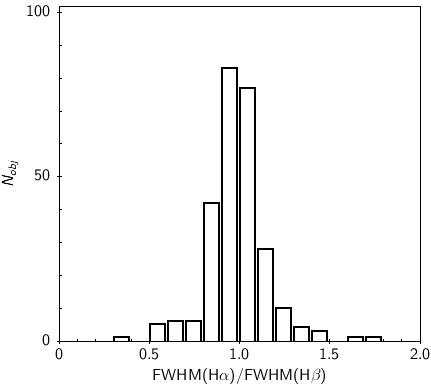}
  \end{center}
  \caption{Left panel: comparison of the H$\beta$\ (blue) and H$\alpha$\ (red) broad line components, after subtraction of the narrow emission lines, plotted in velocity scale and normalized to the same peak intensity in the spectrum of 1RXS~J113247.0+062626. Right panel: distribution of the ratio FWHM(H$\alpha$)/FWHM(H$\beta$) in the broad components of the emission lines from the spectra NLS1 galaxies investigated in \cite{Cracco16}. While the distribution is strongly peaked around 1, favoring the interpretation of an optically thin gas, some differences in the line profiles can arise as a consequence of dust absorption or emission line radiation transfer effects.}
\end{figure}
To further explore the possibility that the presence of ionized gas layers could affect the broad emission lines, we started an investigation of the Balmer line profiles in the broad component of the spectra of a sample of Narrow Line Seyfert 1 galaxies (NLS1s), selected from the SDSS Data Release 7 \cite{Cracco16}. Due to the relatively small width of their broad line components (FWHM$_{{\rm H}\beta} < 2000\, {\rm km\, s}^{-1}$ \cite{Osterbrock85}), the profiles of these lines are little affected by blends with other broad lines, although there are still the narrow components to be accounted for. These, however, are much simpler to model, taking the narrow forbidden lines of [\ion{O}{III}] as templates to constrain their widths and to subtract them from the global profiles. By taking the broad line components in a velocity scale, we are able to compare the resulting profiles, as it is illustrated in Fig.~4. Using a narrow emission line width fixed at 0.75 times the width of [\ion{O}{III}]~$\lambda5007$, in order to account for the larger velocity dispersion of the high ionization gas \cite{Berton15}, we modelled the narrow components of H$\beta$, H$\alpha$\ and the [\ion{N}{II}]~$\lambda\lambda6548,6583$ doublet, trying to isolate the broad emission line profiles. Looking at the resulting FWHM of H$\alpha$\ and H$\beta$, we find that most of the line profiles are very similar, favoring the interpretation of an optically thin gas, though we still observe deviations from this behaviour. It has already been noted that the profiles of these lines can be different in some objects and that they may even exhibit different reverberation lags \cite{Shapovalova10a,Shapovalova10b}. This could point towards a displacement of their emission sites or to a relevant role of dust absorption in the central regions of the source \cite{Gaskell17}, since there is convincing evidence that a substantial amount of dust can exist in the central regions of AGNs, at smaller scales than the NLR \cite{Heard16}.

\section{Discussion}

The processes that occur in the unresolved central regions of AGNs leave characteristic signatures in the emission and absorption components of the observed radiation and they also control which parts of the source are visible. Although a preliminary analysis of the relationships existing between multi-frequency data and optical spectra argue in favor of the Unified Model, a systematic study that collects the huge amount of available observational material still has to be carried out. This type of investigation is highly desirable, due to the invaluable constraints that it could place on the nature and the evolution of the innermost structures in AGN, but it presents obvious difficulties related to the amount of data that should be considered. With our work, we presented a technique to select different types of AGNs, with available spectro-photometric measurements, and we provided some examples on how the comparison of their spectra with the overall SED lead to supporting the unified picture, as well as to the intriguing possibility that SEDs built on the basis of public data could be used to select targets for follow-up spectroscopy or even to attempt preliminary classifications.

A particularly interesting result is the one based on the combination of optical and X-ray spectroscopic observations. In presence of a significant layer of ionized plasma, like the one observed in the X-ray spectrum of PG~1114+445, optical depth effects become important in determining the relative strengths of the recombination lines. In particular they could explain the various emission line intensity ratios that are known to deviate significantly from the standard recombination predictions in Type 1 sources. A substantial increase in optical depth gives raise to two important effects. On the one hand, the H$\alpha$\ photon has a larger absorption cross-section with respect to higher order Balmer photons. On the other, it corresponds to a transition between two adjacent levels and, therefore, has a smaller probability to decay through other spontaneous transition channels, with respect to higher order Balmer photons. Therefore an increase of the line optical depth might lead to a reduction of the high order Balmer line intensities with respect to H$\alpha$, but H$\alpha$ could averagely emerge from an outer layer of the line emitting region. In a dynamical configuration dominated by Keplerian motions, this would imply a narrower line profile. On the other hand, the role played by dust in the emission line regions would possibly lead to the opposite result, suppressing the short-wavelength high-order emission lines more severely than H$\alpha$, therefore favoring an enhancement of this line from deeper regions. Comparing the measurements of the width of the broad emission line profiles, as illustrated in Fig.~4, shows that neither effect is likely to severely affect the emission in the vast majority of sources, although some exceptions may still exist. This result favors the possibility that the optically thin interpretation can be reasonably applied to the broad Balmer lines, suggesting that only a small amount of material should lie along the path of the photons. We can therefore conclude that a thorough investigation of the line intensity ratios along the profiles, which requires data with very high signal to noise ratios, will certainly provide further insight on the not yet completely understood problem fo the BLR structure.

\section{Materials and Methods}

The data and results presented in this paper are based on publicly available services and databases, such as the \emph{VizieR} catalogue service\footnote{\tt http://vizier.u-strasbg.fr/viz-bin/VizieR} and the SDSS \emph{SkyServer}\footnote{\tt http://skyserver.sdss.org/dr14/en/home.aspx}. Plots have been produced with the \emph{TOPCAT}\footnote{\tt http://www.star.bristol.ac.uk/~mbt/topcat/} \cite{Taylor05} and the \emph{XSPEC}\footnote{\tt https://heasarc.gsfc.nasa.gov/xanadu/xspec/} softwares. Most of the illustrated measurements, involving line intensities and photometric colors, were directly extracted from public databases using their standard web interfaces. Exceptions to the illustrated picture are:
\begin{enumerate}
\item measurements of emission line normalized intensities, which were obtained from public SDSS spectra after extracting the BLR contribution by means of multi-Gaussian fits of the observed line profiles performed with {\em IRAF}\footnote{\tt http://iraf.noao.edu/};
\item measurements of the ionized plasma column density, obtained through XSPEC models combining a power-law emission, a soft X-ray thermal excess, an absorption contribution from neutral gas constrained by the \ion{H}{I} column density within the Milky Way and a ionized absorber, applied to a 40~ks observation carried out by {\em XMM Newton} and reduced with the Science Analysis Software (SAS) version 10;
\item the models of SED in multi-frequency AGN data, obtained by appliing Levemberg-Marquardt non-linear $\chi^2$ minimization to combinations of cut-off power laws and black-body contributions.
\end{enumerate}
Table 1 contains a summary of the data sources from which we obtained the SED points, while Table 2 reports a summary of the parameters used in the SED fitting models.

%
%
\vspace{6pt} 

%
\begin{table}[t]
  \caption{Data sources for the selection of multiple wavelength SED points. \label{table:tab01}}
  \centering
  \begin{tabular}{cccc}
  \toprule
      {\bf Instr. / Catalogue} & $\log \nu$ (Hz) & {\bf Band} & {\bf Reference} \\
      \midrule
      NVSS & $9.15$ & Radio & \cite{Condon98} \\
      IRAS & $12.48 -  13.40$ & FIR & \cite{Helou88} \\
      WISE & $13.13 - 13.94$ & FIR & \cite{Cutri12} \\
      2MASS & $14.14 - 14.38$ & NIR & \cite{Skrutskie06} \\
      GALEX & $15.00 - 15.30$ & UV & \cite{Bianchi11} \\
      XMM & $16.86 - 17.68$ & X-rays & \cite{Rosen16} \\
      INTEGRAL & $18.70 - 19.00$ & Soft $\gamma$\ rays & \cite{Bird10} \\
      Fermi/LAT & $23.00 - 26.00$ & $\gamma$\ rays & \cite{Acero15} \\
      \bottomrule
  \end{tabular}
\end{table}

\begin{table}[t]
  \caption{SED fitting models \label{table:tab02}}
  \centering
  \begin{tabular}{lccccccc}
  \toprule
      {\bf Object} & {\bf Function} & $\log \nu_{min}^{(a)}$ & $\log \nu_{max}^{(a)}$ & {\bf Norm.} & {\bf Index}$^{(b)}$ & {\bf Temp.} & $\chi_{red}^2$ \\
      \midrule
      3C 273   & Power law  & $8.6$ & $13.1$ & $2.30 \cdot 10^{-18} $ & $0.60$ & -- & \\
               & Power law  & $13.1$ & $17.6$ & $1.31 \cdot 10^{-6} $ & $-0.31$ & -- & \\
               & Power law  & $17.1$ & $22.6$ & $1.26 \cdot 10^{-23}$ & $0.70$ & -- & \\
               & Black body & -- & -- & $1.03 \cdot 10^{-12}$ & -- & $25800\,$K & $1.11$ \\
      \midrule
      NGC 1068 & Power law  & $8.5$ & $13.0$ & $2.49 \cdot 10^{-28}$ & $1.52$ & -- & \\
               & Power law  & $12.8$ & $17.5$ & $7.068 \cdot 10^2$ & $-0.85$ & -- & \\
               & Power law  & $17.0$ & $24.0$ & $4.82 \cdot 10^{-13}$ & $0.05$ & -- & \\
               & Black body & -- & -- & $1.04 \cdot 10^{-4}$ & -- & $540\,$K & $1.17$ \\
      \bottomrule
  \end{tabular}
  \flushleft{\footnotesize $^{(a)}$ Logarithm of exponential cut-off frequency given in Hz. \\
  $^{(b)}$ The power-law index is given according to the notation $\nu F_\nu \propto \nu^\alpha$.}
\end{table}

\acknowledgments{Funding for this research program has been provided by the Department of Physics and Astronomy of the University of Padua and by the European Space Agency (ESA), under Express Procurement Contract No. 4000111138/14/NL/CB/gp. Additional support from ASTROMUNDUS mobility grants is also gratefully aknowledged.

Funding for the Sloan Digital Sky Survey IV has been provided by the Alfred P. Sloan Foundation, the U.S. Department of Energy Office of Science, and the Participating Institutions. SDSS-IV acknowledges support and resources from the Center for High-Performance Computing at the University of Utah. The SDSS web site is www.sdss.org.

SDSS-IV is managed by the Astrophysical Research Consortium for the Participating Institutions of the SDSS Collaboration including the Brazilian Participation Group, the Carnegie Institution for Science, Carnegie Mellon University, the Chilean Participation Group, the French Participation Group, Harvard-Smithsonian Center for Astrophysics, Instituto de Astrof\'isica de Canarias, The Johns Hopkins University, Kavli Institute for the Physics and Mathematics of the Universe (IPMU) / University of Tokyo, Lawrence Berkeley National Laboratory, Leibniz Institut f\"ur Astrophysik Potsdam (AIP),  Max-Planck-Institut f\"ur Astronomie (MPIA Heidelberg), Max-Planck-Institut f\"ur Astrophysik (MPA Garching), Max-Planck-Institut f\"ur Extraterrestrische Physik (MPE), National Astronomical Observatories of China, New Mexico State University, New York University, University of Notre Dame, Observat\'ario Nacional / MCTI, The Ohio State University, Pennsylvania State University, Shanghai Astronomical Observatory, United Kingdom Participation Group, Universidad Nacional Aut\'onoma de M\'exico, University of Arizona, University of Colorado Boulder, University of Oxford, University of Portsmouth, University of Utah, University of Virginia, University of Washington, University of Wisconsin, Vanderbilt University, and Yale University.

This research has made use of the VizieR catalogue access tool, CDS, Strasbourg, France. The original description of the VizieR service was published in A\&AS 143, 23

This work is based on observations collected with the 1.22m Galileo telescope of the Asiago Astrophysical Observatory, operated by the Department of Physics and Astronomy ”G. Galilei” of the University of Padova.}

\authorcontributions{G.L.M. is the contribution speaker, the main author of the text and the investigator of AGN multi-frequency properties; M.B. adapted the multi-frequency archive with timing capabilities; S.Ch. works on samples of NLS1 galaxies; A.C. contributed to the selection of all emission line spectra with diagnostic lines from SDSS data release 14; S.C. supervised spectroscopic data reduction procedures; E.C. investigated advanced models of NLR emission lines; V.C. selected and analyzed the SDSS DR7 sample of NLS1; M.F. developed models of NLR structure; S.M. worked on multiple Gaussian fits and intensity ratios of optical lines; P.R. conceived the multi-frequency AGN archive and supervised the physical interpretation of optical spectra.}

\conflictsofinterest{The authors declare no conflict of interest. The founding sponsors had no role in the design of the study; in the collection, analyses, or interpretation of data; in the writing of the manuscript, and in the decision to publish the results.} 

\abbreviations{The following abbreviations are used in this manuscript:\\

\noindent 
\begin{tabular}{@{}ll}
AGN & Active Galactic Nucelus\\
BLR & Broad Line Region\\
FWHM & Full Width at Half the Maximum\\
NLR & Narrow Line Region\\
PLTE & Partial Local Thermodynamic Equilibrium\\
SED & Spectral Energy Distribution
\end{tabular}}

\appendixtitles{no} 
\appendixsections{multiple} 
%


\reftitle{References}



\begin{thebibliography}{999}
\bibitem[Acero et al.(2015)]{Acero15}
  Acero, F., Ackermann, M., Ajello, M., et al. Fermi Large Area Telescope Third Source Catalog. {\em ApJS} {\bf 2015}, {\em 218}, id. 23 pp. 41, 10.1088/0067-0049/218/2/23
\bibitem[Antonucci \& Miller(1985)]{Antonucci85}
  Antonucci, R. R. J. \& Miller, J. S. Spectropolarimetry and the nature of NGC 1068. {\em ApJ} {\bf 1985}, {\em 297}, 621-632, 10.1086/163559
\bibitem[Antonucci(1993)]{Antonucci93}
  Antonucci, R. Unified models for active galactic nuclei and quasars. {\em ARA\&A} {\bf 1993}, {\em 31}, 473-521, 10.1146/annurev.aa.31.090193.002353
\bibitem[Baldwin et al.(1981)]{Baldwin81}
  Baldwin, J. A., Phillips, M. M., Terlevich, R. Classification parameters for the emission-line spectra of extragalactic objects. {\em PASP} {\bf 1981}, {\em 93}, 15-19, 10.1086/130766
\bibitem[Berton et al.(2015)]{Berton15}
  Berton, M.; Foschini, L.; Ciroi, S.; Cracco, V.; La Mura, G.; Lister, M. L.; Mathur, S.; Peterson, B. M.; Richards, J. L.; Rafanelli, P. Parent population of flat-spectrum radio-loud narrow-line Seyfert 1 galaxies. {\em A\&A} {\bf 2015}, {\em 578}, id.A28 pp. 12, 10.1051/0004-6361/201525691
\bibitem[Bianchi et al.(2011)]{Bianchi11}
  Bianchi, L., Herald, J., Efremova, B., Girardi, L., Zabot, A., Marigo, P., Conti, A., Shiao, B. GALEX catalogs of UV sources: statistical properties and sample science applications: hot white dwarfs in the Milky Way. {\em Ap\&SS} {\bf 2011}, {\em 335}, 161-169, 10.1007/s10509-010-0581-x
\bibitem[Bird(2010)]{Bird10}
  Bird, A. J., Bazzano, A., Bassani, L., et al. The Fourth IBIS/ISGRI Soft Gamma-ray Survey Catalog. {\em ApJS} {\bf 2010}, {\em 186}, 1-9, 10.1088/0067-0049/186/1/1
\bibitem[Blandford(1986)]{Blandford86}
  Blandford, R. D. Black hole models of quasars. In {\em Quasars}; Swarup, G. \& Kapahi, V. K.; D. Reidel Publishing Co.: Dordrecht, Netherlands, 1986, pp. 359-369, ISBN 90-277-2297-8 (HB), 90-277-2298-6 (PB)
\bibitem[Blanton et al.(2017)]{Blanton17}
  Blanton, M. R., Bershady, M. A., Abolfathi, B., et al. Sloan Digital Sky Survey IV: Mapping the Milky Way, Nearby Galaxies, and the Distant Universe. {\em AJ} {\bf 2017}, {\em 154}, id. 28 pp. 35, 10.3847/1538-3881/aa7567
\bibitem[Condon et al.(1998)]{Condon98}
  Condon, J. J., Cotton, W. D., Greisen, E. W., Yin, Q. F., Perley, R. A., Taylor, G. B., Broderick, J. J. The NRAO VLA Sky Survey. {\em AJ} {\bf 1998}, {\em 115}, 1693-1716, 10.1086/300337
\bibitem[Cracco et al.(2016)]{Cracco16}
  Cracco, V., Ciroi, S., Berton, M., Di Mille, F., Foschini, L., La Mura, G., Rafanelli, P. A spectroscopic analysis of a sample of narrow-line Seyfert 1 galaxies selected from the Sloan Digital Sky Survey. {\em MNRAS} {\bf 2016}, {\em 462}, 1256-1280, 10.1093/mnras/stw1689
\bibitem[Cutri et al.(2012)]{Cutri12}
  Cutri, R. M., Wright, E. L., Conrow, T., et al. Explanatory Supplement to the WISE All-Sky Data Release Products.
\bibitem[Dong et al.(2008)]{Dong08}
  Dong, X., Wang, T., Wang, J., Yuan, W., Zhou, H., Dai, H., Zhang, K. Broad-line Balmer decrements in blue active galactic nuclei. {\em MNRAS} {\bf 2008}, {\em 383}, 581-592, 10.1111/j.1365-2966.2007.12560.x
\bibitem[Elvis et al.(1994)]{Elvis94}
  Elvis, M., Wilkes, B. J., McDowell, J. C., Green, R. F., Bechtold, J., Willner, S. P., Oey, M. S., Polomski, E., Cutri, R. Atlas of quasar energy distributions. {\em ApJS} {\bf 1994}, {\em 95}, 1-68, 10.1086/192093
\bibitem[Gaskell(2017)]{Gaskell17}
  Gaskell, C. M. The case for cases B and C: intrinsic hydrogen line ratios of the broad-line region of active galactic nuclei, reddenings, and accretion disc sizes, {\em MNRAS} {\bf 2017}, {\em 467}, 226-238, 10.1093/mnras/stx094
\bibitem[Heard \& Gaskell(2016)]{Heard16}
  Heard, C. Z. P. \& Gaskell, C. M. The location of the dust causing internal reddening of active galactic nuclei. {\em MNRAS} {\bf 2016}, {\em 461}, 4227-4232, 10.1093/mnras/stw1616
\bibitem[Kewley et al.(2006)]{Kewley06}
  Kewley, L. J., Groves, B., Kauffmann, G., Heckman, T. The host galaxies and classification of active galactic nuclei, {\em MNRAS} {\bf 2006}, {\em 372}, 961-976, 10.1111/j.1365-2966.2006.10859.x
\bibitem[Khachikian \& Weedman(1974)]{Khachikian74}
  Khachikian, E. Y. \& Weedman, D. W. An atlas of Seyfert galaxies. {\em ApJ} {\bf 1974}, {\em 192}, 581-589, 10.1086/153093
\bibitem[Moustakas \& Kennicut(2006)]{Moustakas06}
  Moustakas, J. \& Kennicutt, R. C. Jr. An Integrated Spectrophotometric Survey of Nearby Star-forming Galaxies. {\em ApJS} {\bf 2006}, {\em 164}, 81-98, 10.1086/500971
\bibitem[Osterbrock \& Pogge(1985)]{Osterbrock85}
  Osterbrock, D. E. \& Pogge, R. W. The spectra of narrow-line Seyfert 1 galaxies. {\em ApJ} {\bf 1985}, {\em 297}, 166-176, 10.1086/163513
\bibitem[Popovi\'c(2003)]{Popovic03}
  Popovi\'c, L. \v{C}. Balmer Lines as Diagnostics of Physical Conditions in Active Galactic Nuclei Broad Emission Line Regions. {\em ApJ} {\bf 2003}, {\em 599}, 140-146, 10.1086/376401
\bibitem[Richards et al.(2002)]{Richards02}
  Richards, G. T., Fan, X., Newberg, H. J., et al. Spectroscopic Target Selection in the Sloan Digital Sky Survey: The Quasar Sample. {\em AJ} {\bf 2002}, {\em 123}, 2945-2975, 10.1086/340187
\bibitem[Rosen et al.(2016)]{Rosen16}
  Rosen, S. R., Webb, N. A., Watson, M. G.,et al. The XMM-Newton serendipitous survey. VII. The third XMM-Newton serendipitous source catalogue. {\em A\&A} {\bf 2016}, {\em 590}, id. A1 pp. 22, 10.1051/0004-6361/201526416
\bibitem[Schneider et al.(2010)]{Schneider10}
  Schneider, D. P., Richards, G. T., Hall, P. B., et al. The Sloan Digital Sky Survey Quasar Catalogue. V. Seventh Data Release. {\em AJ} {\bf 2010}, {\em 139}, 2360-2373, 10.1088/0004-6256/139/6/2360
\bibitem[Seyfert(1943)]{Seyfert43}
  Seyfert, C. K. Nuclear Emission in Spiral Nebulae. {\em ApJ} {\bf 1943}, {\em 97}, 28-40, 10.1086/144488
\bibitem[Shapovalova et al.(2010a)]{Shapovalova10a}
  Shapovalova, A. I., Popovi\'c, L. \v{C}., Burenkov, A. N., Chavushyan, V. H., Ili\'c, D., Kollatschny, W., Kova\v{c}evi\'c, A., Bochkarev, N. G., Carrasco, L., Le\'on-Tavares, J., Mercado, A., Valdes, J. R., Vlasuyk, V. V., de La Fuente, E. Spectral optical monitoring of 3C 390.3 in 1995-2007. I. Light curves and flux variation in the continuum and broad lines. {\em A\&A} {\bf 2010}, {\em 517}, id A42 pp. 27, 10.1051/0004-6361/201014118
\bibitem[Shapovalova et al.(2010b)]{Shapovalova10b}
  Shapovalova, A. I., Popovi\'c, L. \v{C}., Burenkov, A. N., Chavushyan, V. H., Ili\'c, D., Kova\v{c}evi\'c, A., Bochkarev, N. G., Le\'on-Tavares, J. Long-term variability of the optical spectra of NGC 4151. II. Evolution of the broad H$\alpha$\ and H$\beta$\ emission-line profiles. {\em A\&A} {\bf 2010}, {\em 509}, id. A106 pp. 20, 10.1051/0004-6361/200912311
\bibitem[Skrutskie et al.(2006)]{Skrutskie06}
  Skrutskie, M. F., Cutri, R. M., Stiening, R., et al. The Two Micron All Sky Survey (2MASS). {\em AJ} {\bf 2006}, {\em 131}, 1163-1183, 10.1086/498708
\bibitem[Taylor(2005)]{Taylor05}
  Taylor, M. B. TOPCAT \& STIL: Starlink Table/VOTable Processing Software. In {\em Astronomical Data Analysis Software and Systems XIV}; Shopbell, P., Britton, M., Ebert, R.; Astronomical Society of the Pacific Conference Series: San Francisco, USA, 2003, 29-33, ISBN 1-58381-215-6
\bibitem[Helou \& Walker(1988)]{Helou88}
  The Joint IRAS Science Working Group, Infrared astronomical satellite (IRAS) catalogs and atlases. Volume 7: The small scale structure catalog; Helou, G. \& Walker, W.; NASA Scientific and Technical Information Division: Washington, DC, 1988, pp. 1-265
\bibitem[Urry \& Padovani(1995)]{Urry95}
  Urry, C. M. \& Padovani, P. Unified Schemes for Radio-Loud Active Galactic Nuclei. {\em PASP} {\bf 1995}, {\em 107}, 803-845, 10.1086/133630
\bibitem[Vaona et al.(2012)]{Vaona12}
  Vaona, L., Ciroi, S., Di Mille, F., Cracco, V., La Mura, G., Rafanelli, P. Spectral properties of the narrow-line region in Seyfert galaxies selected from the SDSS-DR7. {\em MNRAS} {\bf 2012}, {\em 427}, 1266-1283, 10.1111/j.1365-2966.2012.22060.x
\bibitem[Veilleux \& Osterbrock(1987)]{Veilleux87}
  Veilleux, S. \& Osterbrock, D. E. Spectral classification of emission-line galaxies. {\em ApJS} {\bf 1987}, {\em 63}, 295-310, 10.1086/191166
\bibitem[York(2000)]{York00}
  York, D. G., Adelman, J., Anderson, J. E. Jr., et al. The Sloan Digital Sky Survey: Technical Summary. {\em AJ} {\bf 2000}, {\em 120}, 1579-1587, 10.1086/301513
\end{thebibliography}


\end{document}